\documentclass{article}                     
\usepackage{graphicx}
\newcommand{\keywords}[1]{ {\bf Keywords:} #1}
\newcommand{\PACS}[1]{ {\bf PACS:} #1}
\newcommand{\email}[1]{ {email:} #1}

\begin{document}

\title{On the Transits of Solar System Objects in the Forthcoming 
{\sc Planck} Mission: Data Flagging and Coeval Multi-frequency Observations
}


\author{Michele Maris $^1$ and Carlo Burigana $^2$\\
on behalf of the {\sc Planck} Collaboration
 %
 \\
 $^1$
           INAF--OATs, Via Tiepolo 11, I-34143, Trieste, Italy, EU\\
           \email{maris@oats.inaf.it}           \\
           and \\
 $^2$
           INAF--IASF Bologna, Via Gobetti 101, I-40129, Bologna, Italy, EU \\ 
           \email{burigana@iasfbo.inaf.it} 
}

\date{Accepted for the publication to {\em Earth, Moon, and Planets} \\ June 30$^{\mathrm{th}}$, 2009}

\maketitle

\begin{abstract}
 \noindent
In the context of current and future microwave surveys mainly dedicated
to the accurate mapping of 
Cosmic Microwave Background (CMB), 
mm and sub--mm emissions from Solar System will
represent a potential source of contamination as well as 
an opportunity for new Solar System studies.
In particular, the forthcoming ESA {\sc Planck} mission
will be able to observe 
the point-like thermal emission from planets and some large asteroids
as well as the diffused 
Zodiacal Light Emission (ZLE).
After a brief introduction to the field,
we focus on the identification of Solar System discrete objects 
in the {\sc Planck} time ordered data. 
\\ 
\\ 
\end{abstract}

\def\Planck{{\sc Planck}}
\def\deg{^{\circ}}
\def\Tsurf{T_{\mathrm{surf}}}
\def\Teq{T_{\mathrm{eq}}}
\def\Tant{T_{\mathrm{ant}}}
\def\Tantnu{T_{\mathrm{ant},\nu}}
\def\Tb{T_{\mathrm{b}}}
\def\Tbnu{T_{\mathrm{b},\nu}}
\def\radius{R}
\def\range{\Delta}
\def\FWHM{\mathrm{b}}
\def\FWHMnu{\mathrm{b}_\nu}
\def\epsnu{\epsilon_\nu}
\def\radians{\mathrm{radians}}
\def\angdist{\hat{d}}
\def\efficiencynu{K_\nu}
\def\betanu{\beta_\nu}
\def\Dmin{\Delta_{\mathrm{min}}}
\def\Rper{R_{\mathrm{per}}}
\def\Vfov{\dot{\vartheta}_{\mathrm{fov}}}
\def\tvis{{\tau}_{\mathrm{vis}}}
\def\tint{{\tau}_{\mathrm{int}}}
\def\taupoint{{\tau}_{\mathrm{point}}}
\def\tintnu{{\tau}_{\mathrm{int},\nu}}
\def\spinrate{{\omega}_{\mathrm{spin}}}
\def\nspin{{N}_{\mathrm{spin}}}
\def\nobs{{N}_{\mathrm{obs}}}
\def\ndtc{{N}_{\mathrm{dtc}}}
\def\microK{{\mu\mathrm{K}}}
\def\milliK{{\mathrm{mK}}}
\def\thetaspin{{\theta}_{\mathrm{spin}}}

\def\NumberSelected{1221}

\section{Introduction}
\label{intro}

The ESA {\sc Planck} satellite\footnote{http://astro.estec.esa.nl/Planck/}, 
scheduled for launch in May 2009, 
will produce nearly full
sky maps at the frequencies of 30, 44, and
70~GHz (Low Frequency Instrument, LFI; \cite{mandolesi_etal_lfi_98})  
and at 100, 143, 217, 353, 545 and 857~GHz
(High Frequency Instrument, HFI; \cite{puget_etal_hfi_98}) with an
unprecedented resolution (FWHM from
$\simeq33'$ to $\simeq5'$) 
and sensitivity (in the range of $\simeq10-40$~mJy, i.e. $1 - 10$~$\microK$ in terms of antenna temperature,
on a FWHM$^2$ resolution element).
In order to exploit the scientific targets of the mission 
even the contamination from faint foreground sources has to be 
accounted for. External planets,  
Main Belt asteroids (MBAs), and  
diffuse interplanetary dust will be observed
by \Planck\ and will represent a relevant source of
contamination but, at the same time, a good opportunity for 
optical and photometric calibration
(in the case of planets) and an interesting subject of 
scientific analysis\footnote{For a presentation on these topics see e.g. the talk held at the Workshop
 {\em Future Ground based Solar System Research:
Synergies with Space Probes and Space Telescope},
held at Portoferraio, Isola d'Elba, Livorno (Italy), September 8--12, 2008,
with the original title 
{\em Diffuse and Point Like Solar System Emissions in the Forthcoming {\sc Planck} Mission}
(http://www.arcetri.astro.it/$\sim$elba2008; for an account on the conference see also \cite{Messenger}).
}.

The possibility to observe Solar System sources with \Planck\ has 
been assessed in several works.
\cite{BuriPlanets} showed that it will possible 
to use the transits of external planets to reconstruct 
the \Planck\ main beam shapes and to measure the Spectral Energy Distribution 
of the planets.
The expected photometric accuracy is better than $1$\%, being ultimately 
limited by the accuracy of the instrumental calibration
with the cosmological dipole (at least at $\nu \le 353$~GHz).
\cite{Cremonese02} pointed out that \Planck\ will be able 
to observe up to $3\times10^2$ MBAs with a $S/N > 1$ 
and several tens of them with significantly better $S/N$ ratios.
Starting from the COBE model for the Zodiacal Light Emission (ZLE) 
\cite{Kelsall98,FixsenDwek02,Zody06} 
presented a study about the possibility of \Planck\ to 
identify and separate the ZLE
exploiting the seasonal differences in the ZLE signal associated
with the motion of \Planck\ within the Solar System.
Moreover, \cite{Babich07,Babich09} discussed the possibility that the thermal
emission of trans-Neptunian objects (TNOs) and of comets in the Oort
cloud could induce collective distortions in cosmic microwave maps.
At last, \cite{Diego09} analyzed in a general way the problem of how to identify
possible microwave emissions correlated to 
the ecliptical plane
in the context of the large scale anomalies reported in the WMAP maps.

In this work we address how to plan 
observations of Solar System objects
with ground facilities or space satellites at times 
almost coeval with the transits of these sources on \Planck\
receivers, in the view of multifrequency studies of these objects.

\section{Observing the Solar System with the \Planck\ satellite}\label{sec:planck}

\Planck\ is a one--axis stabilized satellite orbiting around the L2 
Earth--Sun libration point \cite{PlanckBlueBook05}.
It is equipped with a 1.5~m effective aperture aplanatic Gregorian telescope the optical 
axis of which points in a direction separated by an angle $\beta = 85\deg$ 
from the spin axis \cite{PlanckBlueBook05}.
During nominal operations the satellite will spin at a rate of 1~r.p.m.
allowing the optical axis of the telescope to scan a
circle on the sky with a semiaperture of $85\deg$ once a minute. 
To survey the whole sky the spin axis orientation will be kept
constant for about 60 minutes, after which it is drifted 
about 2.5~arcmin\footnote{Recently, a constant spacing (of 
2~arcmin), instead of a constant time step, has been agreed.}.
Different ways of drifting the spacecraft spin axis define different scanning strategies
 \cite{PlanckBlueBook05,Dupac05,ScanningSAIT06}.
The spin axis shall never depart more than
a few degrees from the instantaneous antisolar direction.
In the nominal scanning strategy the spin axis is kept on the ecliptic 
plane, while in the baseline scanning strategy currently foreseen 
for \Planck\ it will describe a slow precession about the 
antisolar direction with a semiaperture of $7.5\deg$ and a period of 
six months in order to have complete sky coverage will the whole 
set of receivers \cite{Dupac05}.
The two cryogenic instruments hosted in the telescope focal 
plane, LFI \cite{mandolesi_etal_lfi_98} and HFI \cite{puget_etal_hfi_98}, 
form an array of feed--horns 
covering a field--of--view (f.o.v.) of $\simeq 7\deg$.
Each feed--horn points to a slightly different direction in the sky with 
respect to the center of the f.o.v. defined by the telescope optical axis. 
In most cases the feed--horns of a given frequency are arranged to be 
aligned along the same direction, parallel to the scan circle defined by 
the telescope optical axis, 
as projected on the telescope f.o.v..

The signals collected by each of the feed--horns are detected independently 
by the satellite on--board electronics. The instrument output is thus 
a set of data--streams (time ordered data, TODs)
representing the variation of the antenna temperature (or of 
an equivalent quantity) detected as the telescope scans the sky.
When coupled with accurate information about the instantaneous 
pointing direction of each feed--horn, after proper calibration and
data--processing, the TODs are combined to produce 
multi--frequency maps of the sky (see e.g. \cite{calib_dip,destri_last} 
and references therein).

\Planck\ will return to almost the same heliocentric coordinates every 
12 months but it will complete each sky survey in about 7--8 months.
Because of the value of $\beta$ and the locations of the feed--horns,
\Planck\ will observe Solar System objects in a narrow range of elongations 
($\sim 95\deg \pm 4\deg$) so that the objects of 
the inner Solar System can not be observed.
Sources will be observed by \Planck\ only when they will cross the \Planck\ 
f.o.v. during its scanning of the sky (note that \Planck\ is not a space 
observatory).
There is also a certain intrinsic small uncertainty in predicting 
in advance the exact pointing direction of each feed--horn
as a function of time: a precise time / pointing correlation will be 
established only a posteriori on ground.
Also, the same object will be not observed at the same time by all the feed--horns.

As an example,
we consider the observations of MBAs 
and planets in the case of a mission 
with launch on April 12$^{th}$, 2009 
(JD = 2454935.5)\footnote{This was the planned launch date at 
the date of the Workshop.}. 
In this case the first survey will start about 45~days after the launch 
(May 12$^{th}$, 2009), and it will end 
258~days after the launch
(December 27$^{th}$, 2009).
This marks the begin of the second survey which will end 
471~days after launch
(July 28$^{th}$, 2010).
Since a few details of the current baseline scanning strategy 
are still being defined, we report here results based on the nominal 
scanning strategy. Our results are only 
weakly dependent on the exact scanning strategy.
The simulation has been performed by using the JPL {\em Horizons} 
Ephemeris web 
server\footnote{{\tt http://ssd.jpl.nasa.gov/horizons.cgi}, 
with code for $500@-489$ \Planck.}.
Other missions have developed their own
tools (see e.g. \cite{Barbieri01})
for this type of task.
The \Planck/LFI consortium adopted Horizons
as the reference tool for such kind of calculations for three reasons.
Firstly, Horizons will be the official service by which the detailed 
\Planck\ orbit will be disseminated outside the \Planck\ collaboration.
Horizons allows 
the recovery of the \Planck\ orbit for the nominal launch date and 
to compute the apparent position of any Solar System object 
as seen from \Planck. 
For a given object it is quite simple to estimate the observability
within the \Planck\ f.o.v. by looking at its solar elongation.
Second, our task is  not only to
identify which objects enter the \Planck\ f.o.v. at a given time,
but also to predict precisely their positions with respect to each \Planck\ beam
rotating about the spin axis, 
which can be easily accomplished by using the apparent ecliptical coordinates provided by Horizons.
Third, the number of objects to be tracked is limited
to the five external planets and a relatively small set of 
MBAs 
potentially able to affect \Planck\ data significantly.

The antenna temperature of an asteroid of radius $\radius$ observed at a distance 
$\Delta$ from \Planck\ is computed by using the simple equation \cite{Cremonese02} 

\begin{equation}
  \label{eq:Tant}
\Tantnu \simeq 4 \ln 2 
        \;e^{-4 \ln 2 \angdist^2/\FWHMnu^2} 
        \frac{\efficiencynu }{[\FWHMnu/\radians]^2} 
  \left[\frac{\radius}{\range}\right]^2\Tbnu \, ;
 \end{equation}

\noindent
here $\FWHMnu$ is the FWHM of the feed--horn at the considered frequency 
channel,
$\efficiencynu \le 1$ the photometric efficiency of the considered feed--horn, 
$\Tbnu$ the mean surface brightness temperature of the body,
and $\angdist$ is the angular distance of the object from the feed--horn center.

Since $\Tantnu$ is proportional to $\left[{\radius}/{\range}\right]^2$,
only a small fraction of all of the MBAs are relevant for \Planck.
\cite{Cremonese02} was dedicated to the study of those 
objects observable in at least some \Planck\ feed--horn 
with a $S/N>1$, equivalent to asteroids with 
$\Tantnu$  of several $\microK$, 
corresponding to $\radius/\Delta > 2\times10^{-7}$
and leading to about 300 MBAs observable asteroids.
However, for the accurate exploitation of \Planck\ data 
it is important to take track of
all the foreground sources which could induce 
a contamination 
in one or more of the \Planck\ feed--horns.
To have a significant contamination, a point source should have 
an antenna temperature of at least 
$1\mu\mathrm{K}$.
From eq~(\ref{eq:Tant}) this threshold is 
equivalent to take MBAs having 
$\radius/\Delta > 5\times10^{-8}$ at the epoch of observation.
Since $\Delta$ at the epoch of observation depends on details such as the scanning strategy and 
the exact satellite orbit, we defined our sample of large MBAs by taking all the objects 
having $\radius/\sqrt{\Rper^2-(1.01 \; \mathrm{A.U.})^2} > 5\times10^{-8}$, 
where $\Rper$ is the asteroid mean perihelion distance.

In order to illustrate the observability function 
of asteroids in the \Planck\ f.o.v.,
we use a reference frame
comoving with the spacecraft spin axis having for 
$X$--axis the spin axis, 
the $Y$--axis aligned with the spin axis drift direction,
and the $Z$--axis directed toward the North Ecliptic Pole (NEP).

\begin{figure}
\includegraphics[height=12cm,width=12cm,angle=90]{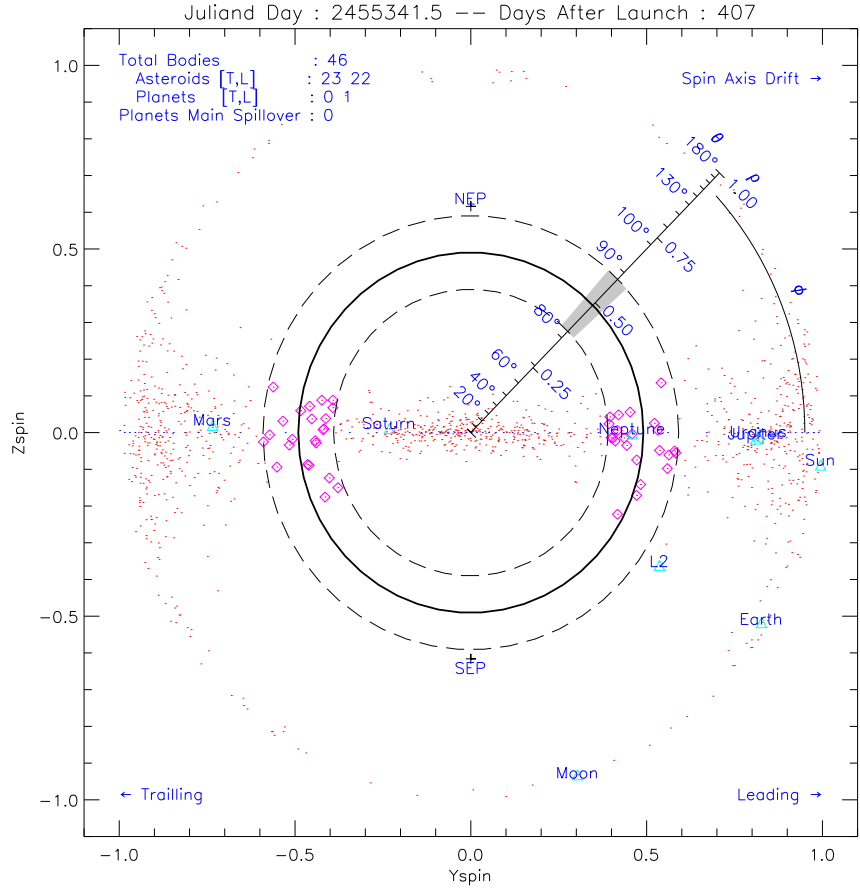}
\caption{
{\sc Planck}--{\em eye} representation of a \Planck\ scan circle 
(concentric rings) an instantaneous \Planck\ f.o.v. (gray region)
together with the positions of a selected subset of asteroids and planets.
The tilted graduated scale in correspondence of the f.o.v. gives the 
relation between the normalized radial coordinate $\rho$ and the angular 
distance from the spin axis $\theta$ according to 
Eq.s~(\ref{eq:normalized:rho:a}--\ref{eq:normalized:rho:c}).
The spin axis points outside the plot at $(0,0)$ and it
drifts toward the positive $Y$ direction defining 
the leading and the trailing directions.
The position of the ecliptic poles (North, NEP and South, SEP) 
are also shown.
Red dots represents asteroids outside the f.o.v. while those transiting 
the f.o.v. are marked with a magenta $\diamondsuit$.
}
\label{fig:eye}       
\end{figure}

The polar plot in Fig.~\ref{fig:eye} represents the position of the 
 \NumberSelected\
MBAs selected on the basis of the above criterion plus all the external planets 
at 00:00~UT of day 407 after the launch.
We define the plotting convention used to drawn the figure 
as the ``{\sc Planck}--{\em eye}'' projection.
The objects are drawn in the $Y,Z$ plane of the spin axis 
reference frame defined above with $\varphi$ being the azimuthal direction of an 
object when projected in the same plane.
The normalized radial coordinate is defined as:

\newcommand{\sign}[1]{\mathrm{sign}[#1]}
\def\eyedump{\varpi}
\def\eyezoom{\zeta}
  \begin{eqnarray}\label{eq:normalized:rho:a}
  t & = & \frac{\theta - \beta }{b_{\mathrm{deg}}}\\
  \label{eq:normalized:rho:b}
  \cal{N} & = &
            \left( 1-e^{-\left({\frac{180\deg - 
\beta}{b_{\mathrm{deg}}} }\right)^\eyedump} \right)^{1/\eyezoom}
           -
            \left( 1-e^{-\left({\frac{\beta}{b_{\mathrm{deg}}} }\right)^\eyedump} 
\right)^{1/\eyezoom}
   \\
   \rho(t) & = & 
\frac{1}{\cal{N}}\left[1+\sign{t}(1-e^{-|t|^\eyedump})^{1/\eyezoom}\right] 
\, , \\
  \sign{t} & = & \left\{ \begin{array}{ll}
  +1, & t \ge 0 \\ -1, &  t < 0 \, \, ,
\end{array}
\right.
 \label{eq:normalized:rho:c}
\end{eqnarray}

\noindent
where $0 \le \rho(\theta) \le 1$, 
$\beta=85\deg$ is the angle between the telescope line of sight and the 
spin axis;
$b_{\mathrm{deg}}$ is a FWHM representative of the instrument; 
$\eyedump$ and $\eyezoom$ are
scaling parameters. Here we take $b_{\mathrm{deg}} = 0.5\deg$, 
$\eyedump=\eyezoom=0.25$.
The scale of the radial coordinate is normalized to have,
$\rho = 0$ for
$\theta=0\deg$, $\rho=1$ for $\theta=180\deg$ and $\rho = 0.5$ for 
$\theta = 85\deg$.
Those scaling parameters are chosen to smoothly ``enlarge'' the sky region 
with $82\deg \le \theta \le 88\deg$
which on a linear scale 
would be too narrow to be seen. Outside that region the relation between
$\rho$ and $\theta$ is compressed and not linear.
In the plot an example of instantaneous location of the rotating f.o.v. 
is represented by the gray rectangle, the circles represents the region 
scanned by the f.o.v.. 
Dots represents the position of the asteroids which are outside of the 
path of the feed--horns, and that consequently are not observed in the
given pointing period. Diamonds the 
position of observable asteroids.
All the objects resides close to  the ecliptic, corresponding to $Y\approx0$. 
Due to the polar singularity, there is an apparent divergence of objects 
near $\rho=1$ (i.e. $\theta=180\deg$).
The spin axis drifts toward the $Y>0$ direction and defines a
leading and a trailing edge of the circle.
At the trailing edge Mars already left the scan circle after having 
transited in it, while Saturn is going to transit.
At the leading edge Neptune is transiting while Jupiter and Uranus are 
going to transit and are overlapped due to the plot scale.
For completeness  the apparent positions of L2, Sun, Moon and Earth, all 
behind the spacecraft, are also marked.

\begin{table}
\caption{Statistics for the transits of the 
Solar System objects of our sample in the \Planck\ f.o.v.
over a 15 months (two surveys) period.
For each quantity, $x$, the table gives its central value
over the selected set of objects and its 
half--range between the members of the set in the form
$\overline{x}\pm\delta x$.
}
\label{tab:0}       
\begin{tabular}{lc}
\hline\noalign{\smallskip}
Total number of Tracked Objects$^\mathrm{a}$: & 1226  \\
Average Number of Tracked Objects in the f.o.v. $^\mathrm{*}$: & $47\pm20$ $^\mathrm{b}$  \\
Fraction of Detectable Tracked Objects : & 31\%  \\
Fraction of Tracked Objects Crossing the f.o.v. once: & 17\%  \\
Fraction of Tracked Objects Crossing the f.o.v. twice: & 83\%  \\
Delay between subsequent transits at leading and at trailing sides: & $164 \pm 6$ days \\
Typical f.o.v. crossing time: & 8--9 days  \\
Apparent f.o.v. crossing speed: & 2 arcmin/hours \\
\noalign{\smallskip}\hline\noalign{\smallskip}
\multicolumn{2}{l}{\small $^\mathrm{a}$ Including planets from Mars to Neptune.}\\
\multicolumn{2}{l}{\small $^\mathrm{*}$ Per pointing period.}\\
\multicolumn{2}{l}{\small $^\mathrm{b}$ The variation refers to difference about minimum and maximum number.}\\
\multicolumn{2}{l}{\small $^\mathrm{c}$ Fraction with respect to the total number of tracked objects.}\\
\noalign{\smallskip}\hline
\end{tabular}
\end{table}

The relative motion of the objects with respect to the telescope
axis scan circle is dominated by the drift
directed toward the $Y>0$ (or leading) edge of the scan circle. 
All the objects enter the leading edge of the scan circle 
and leaves it from the trailing edge more or less moving
parallel to the ecliptic.
An object crossing the scan circle from the right to the left of the plot will 
produce a peak in the data--stream with a one minute period and a 
varying amplitude as it moves through the f.o.v.. 
Each feed--horn will track a well--defined scan circle, relatively close to that 
identified by the telescope axis.
Each object will cross the scan circles corresponding to the various feed--horns
in a sequence and at different times. 


 %
Tab.~\ref{tab:0} reports a set of relevant statistics for the
transits of the Solar System objects of our sample in the \Planck\ f.o.v..
Quantities are presented in the form $\overline{x}\pm\delta x$ where 
$\overline{x}$ is the central value among the subset of bodies 
(which is very near the corresponding average) and 
$\delta x$ is the 
half--range within the sample. 
This should not be interpreted as an error or uncertainty in the prediction.
The table considers a 15 months period i.e. two surveys.
The typical number of objects which may enter the f.o.v. for each 
pointing period is $47 \pm 20$
with an obvious equipartition between the leading and the trailing 
side of each scan circle. 
Among the objects of our sample, about $31\%$ 
are potentially detectable in at least one frequency channel.
This information combined with the number of objects in the f.o.v., 
would lead to an expected rate of detections of about 14 per day.
However, it has to be considered that most of the detections occur at the 
highest HFI frequencies whose corresponding feed horns 
are concentrated in a narrow band near the f.o.v..
This band is crossed by the objects in about one day. 
When this is taken in account the expected number of detectable objects per day reduces to 
$1.6\pm0.6$.
Objects may enter the f.o.v. once or twice during the mission. 
About $17\%$ will enter the f.o.v. once and $83\%$ will enter it twice.
Tracked objects are observed twice if  they cross the \Planck\ f.o.v. 
both at the 
leading and the trailing edge of the scan circle. The typical delay 
between the crossing of the leading and the trailing edge
is $164\pm6$~days.
The typical time for the objects to cross the leading or the 
trailing side of the scan circle
is about 8 or 9 days,
equivalent to say that they cross the f.o.v. at an equivalent speed of about $\Vfov\approx2$~arcmin/hour.
The fact that $\Vfov$ is smaller than the spacecraft
drift rate is due to the relative motion of the objects
with respect to \Planck.
Since the feed--horns are aligned along slightly different scan circles,
the average time in which objects could be observed in their main beam 
is $\tvis\approx\FWHMnu/\Vfov$. 
This ranges from 16~--~17 hours for a 30~GHz beam down to 2~--~3 hours 
for a 857~GHz beam. 
When combined with the number of feed--horns forming each frequency channel, $\ndtc$, 
the total integration time is:

 \begin{equation}
 \label{eq:tint}
  \tintnu\approx\frac{\FWHMnu}{\Vfov\taupoint}
 \frac{\FWHMnu\nspin\ndtc}{\spinrate} \nobs \, ;
\end{equation}

\noindent
here $\nobs$ is the number of observations, $\nspin \approx 55$ the 
number of useful spin rotations during a pointing period
of $\taupoint = 3600$~seconds, and $\spinrate$ is the satellite 
spin--rate.
With the \Planck\ parameters reported in \cite{Cremonese02}, 
$\tintnu$ ranges from 350~sec for the 30~GHz channel
down to 15~sec for the 857~GHz frequency channel,
with a $\pm 40\%$ spread because of detailed FWHM of the 
considered feed--horn, exact beam center angular distance from the spin axis 
direction, and, obviously, object motion.
To a first approximation, useful for practical estimates,

\begin{equation}\tintnu\approx10^{3.94-0.97\log_{10}(\nu_{\mathrm{GHz}})} \; \mathrm{sec}. \end{equation}

 \noindent
As an example of the output of our tool, 
Tab.~\ref{tab:1} gives the epochs for the 
\Planck\ observation of planets and three large asteroids, specifying
also if the object transit in the \Planck\ receiver scan circle occurs 
at its leading or trailing edge. 
Changes in the final orbit \footnote{Included the launch delay from Mid April to Mid March.}, 
as well as spin axis precession, may displace the observation 
windows by up to about a week with respect to the table.

\begin{table}
\caption{Epochs of transits for planets and some large asteroids through the \Planck\ f.o.v.. 
L: transit at the leading edge of the scan circle. T: transit at the trailing edge.
}
\label{tab:1}       
\begin{tabular}{lll}
\hline\noalign{\smallskip}
Mars    & L:2009-Oct-27 $\div$ 2009-Nov-8  & T:2010-Apr-20 $\div$ 2010-May-1  \\
Jupiter & T:2009-Oct-29 $\div$ 2009-Nov-6  & L:2010-Jun-24 $\div$ 2010-Jul-2  \\
Saturn  & L:2009-Dec-26 $\div$ 2010-Jan-2  & T:2010-Jun-10 $\div$ 2010-Jun-17  \\
Uranus  & T:2009-Dec-4 $\div$ 2009-Dec-11  & L:2010-Jun-23 $\div$ 2010-Jun-30  \\
Neptune & T:2009-Nov-4 $\div$ 2009-Nov-12  & L:2010-May-21 $\div$ 2010-May-28 \\
\noalign{\smallskip}\hline\noalign{\smallskip}
Ceres   & L:2010-Mar-24 $\div$ 2010-Apr-1  &                                \\
Pallas  & L:2010-Feb-10 $\div$ 2010-Feb-19 & T:2010-Jul-15 $\div$ 2010-Jul-26 \\
Vesta   & T:2009-Nov-22 $\div$ 2009-Dec-1  & L:2010-May-7 $\div$ 2010-May-16  \\
\noalign{\smallskip}\hline\noalign{\smallskip}
\noalign{\smallskip}\hline
\end{tabular}
\end{table}

\section{Conclusion}

We analyzed the observability
of Solar System discrete objects, such as 
asteroids and planets, 
by the \Planck\ satellite.
This is crucial to plan almost coeval observations
of these sources and to 
flag those \Planck\ data samples which are 
severely affected by Solar System objects.
 %
 %
The main limitation for coeval multifrequency observations 
comes from the fact that \Planck\ 
will only observe each object more or less in quadrature with the Sun. 
Since \Planck\ will keep its spin axis within some degrees from
the antisolar direction, a rough plan for coeval observations can be organized 
by looking at the elongation of the objects with respect to the Sun.
On the other hand, a fine analysis requires
to use on--the--shelf ephemerides calculators
and the available information about
\Planck\ focal plane and scanning strategy.
Our study shows that the majority of the
Solar System objects will be observed at least twice during the nominal 
mission and allows us to provide statistics 
of their transits in the \Planck\ f.o.v.
as well as the corresponding detailed prediction for each object.
MBAs will be observed at an average distance from \Planck\
of about $2.3$~AU so that only asteroids with radii larger than
some tens of kilometers will be observed with a sufficiently high S/N ratio.
On average at least one or two of such objects will be in the \Planck\ f.o.v. each day.

\section*{acknowledgements}
This work has been partially supported by the Italian ASI Contract 
``\Planck\ LFI Activity of Phase E2'' and by the 
``Fondi FFO ricerca libera 2009''. 
We thank Thomas Mueller for constructive discussions. 
The authors acknowledge the anonymous referee for the accurate reading of the test and 
the constructive suggestions which aid at improve considerably the text.



\end{document}